%% file: main.tex
\documentclass[
aps,
prd,
twocolumn,
groupedaddress, 
nofootinbib,
preprintnumbers
]{revtex4-2}

\input{code_stylings}
\usepackage[colorlinks=true, linkcolor=blue]{hyperref}
\usepackage[capitalise]{cleveref}
\usepackage{graphicx}
\usepackage{makecell}
\newcommand{\flare}{\texttt{FLARE} }
\newcommand{\la}{$\langle$}
\newcommand{\ra}{$\rangle$}

\begin{document}
\preprint{ADP-25-23/T1285}

\title{ {\LARGE \flare} \\ \textbf{F}CCee b2\textbf{L}uigi \textbf{A}utomated \textbf{R}econstruction And \textbf{E}vent processing}

\author{C. Harris\footnotemark[1]}
\author{A. Desai\footnotemark[2]}
\footnotetext[1]{Email: {\color{blue}cameron.harris@adelaide.edu.au}}
\footnotetext[2]{Email: {\color{blue}aman.desai@adelaide.edu.au}}

\affiliation{Department of Physics, The University of Adelaide, North Terrace, SA 5005}

\date{\today}

\begin{abstract}
\flare is an open source data workflow orchestration tool designed for the FCC Analysis software and \texttt{Key4HEP} stack. Powered by \texttt{b2luigi}, \flare automates and orchestrates the \texttt{fccanalysis} stages from start to finish. Furthermore, \flare is capable of managing the Monte Carlo (MC) data workflow using generators inside the \texttt{Key4HEP} stack such as \texttt{Whizard}, \texttt{MadGraph5\_aMC@NLO}, \texttt{Pythia8} and \texttt{Delphes}. In this paper the \flare v0.1.4 package will be explored along with its extensible capabilities and a feature rich work environment. Examples of \flare will be discussed in a variety of use-cases, all of which can be found at \href{https://github.com/CamCoop1/FLARE-examples/}{https://github.com/CamCoop1/FLARE-examples}. The open source repository of \flare can be found at \href{https://github.com/CamCoop1/FLARE}{https://github.com/CamCoop1/FLARE}
\end{abstract}

\maketitle

\section{Introduction} 

The Future Circular Collider is a proposed particle collider to be operated at the European Organization for Nuclear Research~\cite{FCC:2025lpp,FCC:2025uan,FCC:2025jtd}. It is planned to start operations in the 2040's. This collider is planned to first run as an $e^+e^-$ operating at energies $\sqrt{s} = 88 $ GeV up to $365$ GeV. This collider will present physics opportunities in Electroweak Physics, Higgs Physics, Quantum Chromodynamics Physics, and Beyond the Standard Model physics~\cite{Selvaggi:2025kmd, dEnterria:2025hbe, Azzi:2025dwl}. The measurement of physics observables would rely on the performance of the detectors. For the $e^+e^-$ collider various detector designs have been proposed~\cite{FCC:2025lpp}. In order to assess the merits of different detector designs, and to optimize the detector technologies to carry out precision Standard Model measurements, several physics studies are carried out implementing Monte Carlo Simulation. These studies are implemented within the \texttt{Key4HEP} software stack~\cite{Key4hep:2023nmr} and using FCCAnalyses software~\cite{helsens_2025_15528870}. The \texttt{Key4HEP} software was developed as a way to allow different experimental collaborations to perform studies on proposed future colliders. These studies require Monte Carlo tools so as to generate hard scattering matrix elements, simulate parton shower and to incorporate detector effects. All of the necessary tools to fulfill these requirements can be sourced from the single umbrella package, \texttt{Key4HEP}. The FCCAnalyses software was developed to aid in performing studies of the proposed FCC experiment, using Monte Carlo produced via the \texttt{Key4HEP} stack. The \texttt{Key4HEP} and FCCAnalyses software are excellent in their capabilites however, there is not yet a clear way to harmoniously synchronize the two packages.

Our package, \flare,  works on top of \texttt{Key4HEP} stack and FCCAnalyses software. In particular, the package has been designed to be able to generate samples with \texttt{MadGraph5\_aMC@NLO}~\cite{Alwall_2014}, Whizard~\cite{Kilian_2011}, Pythia~\cite{Sjostrand:2006za,Sjostrand:2014zea}, \texttt{Delphes}~\cite{deFavereau:2013fsa}, and then also enable analysis on the generated events. In affect coordinating the two separate packages, \texttt{Key4HEP} and FCCAnalyses in the background, automatically whilst giving the user an easy interface to work within. This is possible by use of the python package \texttt{b2luigi}~\cite{b2luigi2025}, a workflow orchestration tool managed by the Belle II Collaboration. \flare leverages \texttt{b2luigi} and its powerful orchestration tooling to create an abstracted layer on top of \texttt{Key4HEP} and FCCAnalyses syncronizing the two softwares. This allows for a wide range of analyses to be performed easily.  \flare has been designed with extensibility in mind, ensuring that other tooling similar to \texttt{Key4HEP} and FCCAalyses can be added later and coordinate with the existing software implemented inside \flare.

In \Cref{sec:flare-package} we discuss the \flare package as a whole, from installation in \Cref{subsec:installation}, FCCAnalyses configuration in \Cref{subsec:Analysis}, MC Production configuration in \Cref{subsec:mcproduction-config}, adding more workflows in \Cref{subsec:adding-workflows}, the \flare Command Line Interface tool in \Cref{subsec:cli} and lastly, custom workflows within the \flare environment inside \Cref{subsec:custom-flare-workflow}.

In \Cref{sec:flare-examples} we display various example analyses using \flare. Firstly in \Cref{sec:higgs-mass} we show the first example, replicating the Fast Simulation Higgs Mass example from the \href{https://github.com/HEP-FCC/FCCAnalyses/tree/master/examples/FCCee/higgs/mH-recoil}{FCCee higgs mH-recoil example}\footnote{FCCee higgs mH-recoil example - https://github.com/HEP-FCC/FCCAnalyses/tree/master/examples/FCCee/higgs/mH-recoil}. In \Cref{sec:example-large-batch-time} we conduct a timed study into the MC Production capabilities of \flare, generating MC for four $e^+e^-$ decays. \Cref{sec:example-whizard-crossection} displays \flare's flexibility showing how an analyst can orchestrate their own \texttt{b2luigi} workflows using the internal packaged \flare \texttt{b2luigi} tasks. The example shows a workflow to calculate the cross section of a decay chain $e^+e^-$ and compares to centrally produced MC samples from the FCC Collaboration as a control set. Lastly \Cref{sec:example-full-fastsim} brings the many features of \flare to the forefront and extends the example from \Cref{sec:higgs-mass}. This time scheduling the MC production for a single signal sample using multiple detector configurations. These output rootfiles are then analysed by the Fast Simulation Higgs analysis and we investigate key variables and how they vary depending on the detector card used. 

\section{The \flare package: v0.1.4 \label{sec:flare-package}}

The \flare package has been created to automate and streamline any FCC analysis which may or may-not require its own Monte Carlo production. The framework has been developed with ease of use in mind and requires very little configuration from the user. The framework leverages the powerful python package \texttt{b2luigi} to manage and run the complex workflows that are commonplace for HEP analyses in general. \flare can be easily extended to orchestrate and interface different independent packages together. \flare is open sourced and can be found here \href{https://github.com/CamCoop1/FLARE}{https://github.com/CamCoop1/FLARE}

\subsection{\texttt{b2luigi} Overview \label{subsec:b2luigi}}
\texttt{b2luigi} is a package maintained by the Belle II experiment and is an extension of the luigi \cite{luigi2025} package created by Spotify. In simplest terms, ``\textit{it helps you schedule working packages (so-called tasks) locally or on a batch system. Apart from the very powerful dependency management system by luigi, \texttt{b2luigi} extends the user interface and has a built-in support for the queue systems, e.g. LSF and HTCondor}" \cite{b2luigi2025}. By packaging small portions of code as tasks, \texttt{b2luigi} can orchestrate the scheduling and running of these tasks allowing an easy interface to create and manage highly complex workflows commonly seen in High Energy Particle Physics (HEP). \texttt{b2luigi} uses Directed Acyclic Graphs (DAG) to orchestrate workflows of any complexity~\cite{b2luigi2025}. 

Computing clusters are common place in HEP. Often these clusters manage their computing resources via batch systems, such as HTCondor \cite{condor-practice}. This allows for optimisation and scheduling of compute resources for a large user base. \texttt{b2luigi} allows users to create tasks that are intended to run on such batch systems. To tell \texttt{b2luigi} what batch system one wishes to submit to, one must set the \texttt{batch\_system} setting. This setting can be set at the Task level by attaching a \texttt{batch\_system} property to the Task class with the corresponding batch system to be used, as shown below. 

\begin{lstlisting}[language=python]
# /user/bin/python
import b2lugi as luigi 

class SlurmBatchTask(luigi.DispatchableTask):
    """
    Basic implementation of HTCondor batch task
    """
    batch_system = `slurm'

    def process(self):
        ...

    def output(self):
        ...
\end{lstlisting}

 If a user needs to run their workflow on a different batch system, one need only change this setting. For completeness at the time of publication, \texttt{b2luigi} version 1.2.2~\cite{b2luigi2025} supports the following batch systems:
 
\begin{itemize}
    \item LSF \cite{IDM-lsf}
    \item HTCondor \cite{condor-practice}
    \item Gbasf2 (internal Belle II grid) \cite{gbasf2}
    \item Apptainer \cite{apptainer}
    \item Slurm \cite{slurm}
\end{itemize}

For more information on b2lugi, refer to their documentation \href{https://b2luigi.belle2.org/}{https://b2luigi.belle2.org/}.

\subsection{Software Installation \label{subsec:installation}}
To access the \flare software, you can simply use pip (or any python package manager) to install it from PyPI, the link to which can be found here \href{https://pypi.org/project/hep-flare/}{hep-flare}\footnote{https://pypi.org/project/hep-flare/}. Running the command below in a terminal will install \flare using pip.
\begin{lstlisting}
pip3 install hep-flare\end{lstlisting}

To ensure \flare works as intended, one should first setup the latest \texttt{Key4HEP} stack from the CVMFS and create a virtual environment in which to install \flare into. 

\begin{lstlisting}
$ source /cvmfs/fcc.cern.ch/sw/latest/setup.sh
$ python3 -m venv .venv
$ source .venv/bin/activate
(.venv)$ pip3 install hep-flare\end{lstlisting}

\subsection{Configuration for FCC Analysis Workflow \label{subsec:Analysis}}
The analysis portion of this framework uses the \texttt{fccanalysis} package available in the CernVM File System (cvmfs). The \texttt{fccanalysis} commandline interface accommodates 3 ordered stages of data manipulation and analysis, each requiring an accompanying Python script. 

\begin{itemize}
    \item \texttt{fccanalysis \textcolor{red}{run} run\_script.py }
    \item \texttt{fccanalysis \textcolor{red}{final} final\_script.py }
    \item \texttt{fccanalysis \textcolor{red}{plots} plots\_script.py }
\end{itemize}
It is common place for the \texttt{run} option to be used twice, refered to as Stage1 and Stage2. Generally Stage1 builds tracks from charged particles and basic final state decay products. Stage2 reconstructs analysis specific parent particles using the final state particles and other detector information. The \texttt{final} stage creates ROOT histograms of variables declared by the analyst. The \texttt{plots} stage allows an analyst to create formatted plots using the \texttt{final} or Stage2 output, depending on preference. 

\smallbreak

Each of these stages have been encased into their own \texttt{b2luigi} task and allow a user to define their own workflow by simply adding their python scripts into the current working directory. To specify the workflow required for a given analysis, one must prefix each python script with its corresponding stage e.g:

\begin{lstlisting}
$ ls analysis/
$ stage1.py final.py plots.py
\end{lstlisting}

This tells \flare that it must schedule \texttt{fccanalysis} tasks for Stage1, \texttt{final}, and \texttt{plots}. A user may create any ordered analysis workflow they please using this simple prefix rule. The following are the available ordered prefixes one can use for an FCC Analysis inside \flare v0.1.4

\begin{itemize}
    \item stage1
    \item stage2
    \item final
    \item plots
\end{itemize}

\subsection{Monte Carlo Production Configuration \label{subsec:mcproduction-config}}
\flare also makes scheduling of Monte Carlo (MC) production easy. The MC generators detailed below have been packaged into \texttt{b2luigi} Tasks.

\begin{itemize}
    \item Whizard
    \item \texttt{MadGraph5\_aMC@NLO}
    \item \texttt{Pythia6}
    \item \texttt{Pythia8}
\end{itemize}

Users can access these MC generators by configuring \flare accordingly. The MC configuration of \flare is handled by a single YAML file in which you define all the required settings such as the generator, data types, batch submission environments etc. As of \flare v0.1.4 this YAML file must also contain a \texttt{"$model"="UserMCProdConfigModel"$} variable, which tells \flare how to validate the input configuration to ensure all required information is present. The YAML file must be located in a folder in your current working directory called \texttt{mc\_production}. The name of the YAML file is arbitrary, \flare will search for any YAML file present inside the \texttt{mc\_production} directory. For example, if the Whizard generator was required to create $e^+e^-\to Z\bar{Z}$ one would create a configuration YAML file like so:
\begin{lstlisting}
# mc_production/details.yaml
"$model" : "UserMCProdConfigModel"

global_prodtype : whizard

datatype:
    - wzp6_ee_ZH_ecm240\end{lstlisting}

Depending on the generator that is required, the accompanying input files must be added by the analyst into the \texttt{mc\_production} folder. 

\subsubsection{Whizard + DelphesPythia6 \label{subsubsec:whizard_delphes6}}
To select the Whizard + DelphesPythia6 workflow, one must set the \texttt{global\_prodtype = whizard} inside the YAML file. Under \texttt{datatype} one must list all the datatypes to be generated. The exact names of each datatype is for the choosing of the analyst. To run the Whizard + DelphesPythia6 workflow the following files must be located in the \texttt{mc\_production} directory:

\begin{itemize}
    \item ++.sin
    \item card\_\la \ra.tcl
    \item edm4hep\_\la \ra.tcl
\end{itemize}

here \la\ra\space indicates areas where a user can input their own naming conventions. \flare checks for the key words and suffixes around the symbolic place holders ++ an \la\ra. The ++ placeholder indicates that there must be a .sin file for each datatype, e.g if a datatype inside the configuration YAML is named \texttt{wzp6\_ee\_nunuH\_Hbb\_ecm240} one would have to name the corresponding .sin file as \texttt{wzp6\_ee\_nunuH\_Hbb\_ecm240.sin}. Each .sin file must have its output file named proc (the standard inside FCC). If not the software will not be able to work correctly.

\subsubsection{\texttt{MadGraph5\_aMC@NLO} + DelphesPythia8 \label{subsubsec:madgraph_delphes}}
To select the \texttt{MadGraph5\_aMC@NLO} + DelphesPythia8 workflow, one must set the \texttt{global\_prodtype = madgraph} inside the MC production configuration YAML file. Under \texttt{datatype} one must list all the data types to be generated. The exact names of each data type is for the choosing of the analyst. To run the \texttt{MadGraph5\_aMC@NLO} + DelphesPythia8 workflow the following files must be located in the \texttt{mc\_production} directory:

\begin{itemize}
    \item ++\_runcard.dat
    \item card\_\la\ra.tcl
    \item edm4hep\_\la\ra.tcl
    \item pythia\_card\_\la\ra.cmd
\end{itemize}
    
here \la\ra \space indicates areas where an analyst can input their own naming conventions. \flare checks for the key words and suffixes around the symbolic placeholders ++ and \la\ra. The ++ placeholder indicates there must be a .dat file for each datatype declared inside the configuration YAML analagous to \Cref{subsubsec:whizard_delphes6}. The pythia\_card\_\la\ra.cmd file must have the variable \texttt{Beams:LHEF = signal.lhe}. If this is not present, the software will be unable to run the DelphesPythia8\_EDM4HEP command.

\subsubsection{DelphesPythia8}
To select just the DelphesPythia8 workflow, one must set the \texttt{global\_prodtype = pythia} inside the MC Production configuraion YAML file. In the \texttt{datatype} list one must list all the datatypes to be generated with identifiable names. The exact names of each dataype is for the choosing of the analyst. To run the DelphesPythia8 workflow the following files must be located in the \texttt{mc\_production} directory
\begin{itemize}
    \item card\_\la\ra.tcl
    \item edm4hep\_\la\ra.tcl
    \item ++.cmd
\end{itemize}
    
here \la\ra\space indicates areas where a user can input their own naming conventions. \flare checks for the key words and suffixes around the symbolic placeholders ++ and \la\ra. The ++ placeholders indicates there must be a .cmd file for each datatype listed inside the configuration YAML.

\subsubsection{Specific Environment for Batch Jobs }
\label{subsubsec:specific-env}

If a specific environment is required to be setup for each MC Production batch job one can parse a value to the \texttt{global\_env\_script\_path} variable inside the MC Production configuration YAML. This value can be the setup script filename which is assumed to be inside the \texttt{mc\_production} directory. If instead a setup script is located outside the \texttt{mc\_production} directory, an absolute path can be given. An example MC Production configuration YAML is shown below. 

\begin{lstlisting}
# MC Production config.yaml
"$model" : "UserMCProdConfigModel"

global_prodtype: pythia8

global_env_script_path: setup.sh

datatype:
    - p8_ee_ZZ_ecm240 
    - p8_ee_ZH_ecm240 
\end{lstlisting}

\subsubsection{Multiple Detector Card Production}
\label{subsub:multi-card}
 
\flare is capable of generating MC using multiple different detector cards. Given the MC Production configuration file shown below, \flare will take the product of the \texttt{datatype} list with the \texttt{card} list.

\begin{lstlisting}
# MC Production config.yaml
"$model" : "UserMCProdConfigModel"

global_prodtype: pythia8

datatype:
    - p8_ee_ZZ_ecm240 
    - p8_ee_ZH_ecm240 
card:
    - card_IDEA
    - card_IDEA_lighterVXD_35pc
    - card_IDEA_lighterVXD_50pc
\end{lstlisting}

\begin{enumerate}
    \item \texttt{p8\_ee\_ZZ\_ecm240 * card\_IDEA}
    \item \texttt{p8\_ee\_ZZ\_ecm240 * card\_IDEA\_lighterVXD\_35pc}
    \item \texttt{p8\_ee\_ZZ\_ecm240 * card\_IDEA\_lighterVXD\_50pc}
    \item \texttt{p8\_ee\_ZH\_ecm240 * card\_IDEA}
    \item \texttt{p8\_ee\_ZH\_ecm240 * card\_IDEA\_lighterVXD\_35pc}
    \item \texttt{p8\_ee\_ZH\_ecm240 * card\_IDEA\_lighterVXD\_50pc}
\end{enumerate}

In this demonstration, the MC Production workflow would output six root files. The naming follows a conventional format \texttt{<datatype>\_<card>.root} to easily identify what each file corresponds to e.g, \texttt{p8\_ee\_ZZ\_ecm240\_card\_IDEA\_lighterVXD\_35pc.root}

\subsubsection{Multiple Production Types}
Inside of \flare it is possible to produce any number of MC types, each with their own specific generator. For this, the \texttt{global\_prodtype} is not needed in the configuration YAML instead each data type in the list specifies its own local \texttt{prodtype} value. 

\begin{lstlisting}
# MC Production config.yaml
"$model" : "UserMCProdConfigModel"

datatype:
    - p8_ee_ZZ_ecm240:
        prodtype : pythia8
    - wzp6_ee_nunuH_Hbb_ecm240:
        prodtype : whizard
\end{lstlisting}

If a configuration YAML like the one above is passed to \flare it will submit in parallel to the batch system each \texttt{datatype} using its specified \texttt{prodtype}.

\subsection{Adding More ``\flare Workflows" \label{subsec:adding-workflows}}

As of \flare v0.1.4 there are only two main Workflows, the FCC Analysis and select MC Production generators from the \texttt{Key4HEP} stack. However, extendability has been at the center of \flare and its development.  \flare has been designed to easily add more workflows and extend existing ones such as adding new MC Production generators from \texttt{Key4HEP} as required. This is achieved by following a process heavily inspired by github actions which we call ``\flare Workflows". \flare is able to build entire workflows from a single YAML file which details all the information required for any number of ordered stages. For this automatic workflow generator to work within \flare, a system called `Bracket Mappings' was developed that allows a developer to declare precisely what format each argument of the command line executable should be, whether it should be determined by the user or \flare and so on. This is done through the use of symbolic placeholders that during runtime are replaced with the appropriate string data.

As an example, if one wishes to use the Whizard command line tool to create a \texttt{wzp6\_ee\_nunuH\_Hbb\_ecm240} sample, the executable would be formatted as below.

\begin{lstlisting}
$ whizard sample.sin \end{lstlisting}

where the \texttt{sample.sin} is configured for $e^+e^-\to H\nu\bar{\nu}\to b\bar{b}\nu\bar{\nu}$ generation. Given the output is set correctly to be \textbf{proc} this command line executable would generate a \texttt{proc.stdhep} file in its output. Translating that into a \flare Workflow gives the following:

\begin{lstlisting}
whizard:
    stage1:
        cmd : whizard {0} 
        args :
            - sample.sin
        output_file : proc.stdhep\end{lstlisting}
where in the most basic form, a \flare Workflow requires a unique identifier as the top level value (in this case ``whizard"). Below this top level value is the ordered stages of the Workflow, in this simple example there is only one stage, hence \texttt{stage1}. Each stage requires three fields to be entered below it:

\begin{itemize}
    \item \textbf{cmd}: the command line interface entry point along with ordered fields which \flare will fill with the \texttt{args}
    \item \textbf{args}: the ordered list of arguments which the command line interface requires as per its formatting
    \item \textbf{output\_file}: the name of the required output file from this workflow
\end{itemize}

\flare will then build a \texttt{b2luigi} task using all this information, submitting a \texttt{subprocess} call with the provided \texttt{cmd} and ordered \texttt{args} and is only completed once the \texttt{output\_file} is detected. In the naive example above, the case where more than one MC sample type is to be generated is not covered since one cannot have multiple files called \texttt{sample.sin}. Instead, in production, \flare uses one of the Bracket Mappings to specify the format of the .sin file, for example:

\begin{lstlisting}
whizard:
    stage1:
        cmd : whizard {0}
        args :
            - ++.sin
        output_file : proc.stdhep\end{lstlisting}
where the only element of the \texttt{args} list \texttt{++.sin} is telling \flare that the name of the .sin file needs to match one of the data types declared in the MC Production configuration file, as described in \Cref{subsubsec:whizard_delphes6}. This is just one of the five Bracket Mappings available inside \flare Workflows to allow much more flexibility with the formatting of input files. The five Bracket Mappings are detailed in \Cref{tab:bracket_mappings}. The production \flare Workflow of the \texttt{whizard} MC generator is shown below and displays all five Bracket Mappings being used.

\begin{lstlisting}
whizard :
  stage1 :
    cmd : whizard {0}
    args :
      - ++.sin
    output_file: proc.stdhep
  stage2:
    cmd : DelphesSTDHEP_EDM4HEP {0} {1} {2} {3}
    args:
      - card_<>.tcl
      - edm4hep_<>.tcl
      - ().root
      - --.stdhep
    output_file : $$.root\end{lstlisting}
Note in \Cref{subsubsec:whizard_delphes6} it was discussed that the user must provide a \texttt{++.sin} file for each data type, a \texttt{card\_\la\ra .tcl} file and a \texttt{edm4hep\_\la\ra .tcl} file where the user may add their own naming conventions within the \la\ra\space brackets. \flare Workflows in conjunction with Bracket Mappings is precisely why these requirements are in place. The FCC stages Workflow described in \Cref{subsec:Analysis} also use \flare Workflows and Bracket Mappings to define \texttt{fccanalysis} within \flare. 

\begin{table*}[]
    \centering
    \caption{Bracket mappings used by \flare to interpret file names and output locations as of \flare version 0.1.4. Bracket Mappings is a symbolic placeholder language used by \flare}
    \begin{tabular}{|c|p{0.75\textwidth}|}
    \hline
        \textbf{Bracket Mapping} &  \textbf{Description} \\ \hline\hline
         \la\ra & Tells \flare that the user may name this file whatever they like but that the prefix/suffix must be formatted accordingly. For example, given \texttt{card\_\la\ra.tcl} one could name this file \texttt{card\_hello\_world.tcl} and \flare will still locate the file.  \\ \hline
         () & Tells \flare that it needs to fill this field with the output path from the \texttt{b2luigi} task for this given stage. This must \textcolor{red}{always} be done when the output file name is part of the input arguments for the command line tool. \\ \hline
         - - & Tells \flare to fill this field with the path to the output file from the previous stage. In \texttt{b2luigi} this is commonly referred to as the `input path' of a Task. \\ \hline
         ++ & Tells \flare to match this placeholder with the provided data types inside the MC Production config YAML \\ \hline
         \$\$ & Mainly used for output file naming when \flare can dynamically adjust the output file name depending on the input parameters of the associated \texttt{b2luigi} task. For example, when running the MC Production workflow with multiple detector cards, \flare will adjust the name of each output rootfile to reflect not only the data type produced by the card that was used. \\ \hline
    \end{tabular}
    \label{tab:bracket_mappings}
\end{table*}

\subsection{\flare Command Line Interface \label{subsec:cli}}
In this section the \flare Command Line Interface (CLI) is detailed along with how to use it to configure an FCC analysis or MC Production Workflow. After installing \flare (see \Cref{subsec:installation}) the CLI will be available in the terminal. The entry point for the CLI is the keyword \texttt{flare}. To access the help documentation for the CLI run the following command:

\begin{lstlisting}[style=plain]
$ flare --help
usage: flare [-h] {run} ...

CLI for FLARE Project

positional arguments:
  {run}
    run       Run the flare command

options:
  -h, --help  show this help message and exit\end{lstlisting}
At the time of publication, the only command in the \flare CLI v0.1.4 is \texttt{run} however, this redundancy leaves room for future features to be added to the CLI. Accessing the help documentation for the \texttt{run} command yields the following.

\begin{lstlisting}[style=plain]
$ flare run --help
usage: flare run [-h] {analysis,mcproduction} ...

positional arguments:
  {analysis,mcproduction}
    analysis     
        Run the FCC analysis workflow
    mcproduction 
        Run the MC Production workflow\end{lstlisting}
The \texttt{run} command has two sub-commands being \texttt{analysis} and \texttt{mcproduction}. The exact use cases of these commands will be discussed in later sections however, the configuration of each sub-command is identical. Investigating the \texttt{analysis} help documentation yields the following:

\begin{lstlisting}[style=plain]
$ flare run analysis --help
usage: flare run analysis [-h] [--name NAME] [--version VERSION] [--description DESCRIPTION] [--study-dir STUDY_DIR] [--output-dir OUTPUT_DIR] [--config-yaml CONFIG_YAML] [--mcprod]

options:
  -h, --help            
            show this help message and exit
  --name NAME
            Name of the study
  --version VERSION
            Version of the study
  --description DESCRIPTION
            Description of the study
  --study-dir STUDY_DIR
            Path to directory of input files, default is current working directory
  --output-dir OUTPUT_DIR
            Path to outputs will be produced, default is the current working directory
  --config-yaml CONFIG_YAML
            Path to a YAML config file contained in STUDY_DIR
  --mcprod  
            If set, also run mcproduction as part of the analysis\end{lstlisting}

Each option in the \flare CLI has default behaviour, meaning it is to the analyst discretion how much customization they wish to have when using \flare. Each option and its default behaviour is detailed in \Cref{tab:cli-information}. It is important to note that \flare uses the options \textit{name} and \textit{version} when building the result directories in which all \flare outputs will be stored, formated like \texttt{\$CWD/<name>/<version>}. If an analyst makes a change to their code and but wishes to keep their outputs in a neat ordered fashion, consider changing the \textit{version} option.

\begin{table*}[]
    \centering
    \caption{\flare CLI options, their defaults and descriptions of functionality as of \flare v0.1.4.}
    \begin{tabular}{|c|p{0.16\textwidth}|p{0.74\textwidth}|}
         \hline \textbf{CLI Option} & \textbf{Default} & \textbf{Description}  \\ \hline \hline
         name & ``default\_name" & The name of the analysis being conducted \\ \hline 
         version & ``1.0" & The version of the analysis. This can be whatever identifiable name the users wishes, not just ordered numbers \\ \hline 
          description & NONE & A description of the analysis being conducted, this is saved as a .txt file in the results directory for book keeping purposes  \\ \hline
          study-dir & Current working directory & The path to the directory in which the analysis input files are being kept. This is most useful when multiple studies are being conducted in an analysis requiring different input files. By default the CURRENT WORKING DIRECTORY is used \\ \hline
          output-dir & Current working directory & The path to the directory where the results of the \flare workflow will be kept. By default the CURRENT WORKING DIRECTORY is used \\ \hline 
          config-yaml & a YAML in the current working directory & relative or absolute path to the  \underline{directory} containing the config YAML. Used as a way to neatly store all CLI options in one place along with additional \texttt{b2luigi} settings, reducing the need to repeatedly type all required options as CLI arguments \\ \hline
          mc-prod & \makecell[l]{\textbf{\underline{Sub-Command}} \\ \texttt{analysis} : False \\ \texttt{mcproduction}: True} & \textcolor{red}{Only available for \texttt{analysis} CLI sub-command, \texttt{mcproduction} always sets mc-prod=True}. If True, set the MC Production workflow as a requirement before the FCC Analysis workflow, informing \flare that the outputs of the MC Production workflow are the inputs for the FCC Analysis. \\ \hline
    \end{tabular}
    \label{tab:cli-information}
\end{table*}

If one wishes to simplify their use of the \flare CLI tool, a YAML file containing the values of all required CLI options can be created. The naming of the YAML is arbitrary and is the users choice. This YAML file also allows the user to pass any \texttt{b2luigi} specific settings into \flare for use during runtime. For example:

\begin{lstlisting}[style=plain]
# FLARE Setting
name : paper_examples
version:  higgs_mass/run1
description: Paper examples of MC production, analysis workflow
studydir: analysis/studies/higgs_mass_example

# b2luigi settings
batch_system: slurm\end{lstlisting}
Here we note that there are two distinct sets of settings, one for \flare CLI options and one for \texttt{b2luigi}. If no \texttt{batch\_system} is given by the user, \flare will set a default value ``local" which tells \texttt{b2luigi} to not submit to any batch system, rather run the entire workflow on a single process. Refer to \Cref{subsec:b2luigi} for all available batch systems and further \texttt{b2luigi} information. By default \flare will always look for a configuration YAML file in the current working directory, even when one is not in use. However, by parsing to the \texttt{config-yaml} CLI option a relative path to a directory containing the YAML, \flare will find and load those settings. If one wants to be specific a full path to the YAML file can be passed.

It is important to note that \flare has a hierarchy when it comes to setting these CLI options. \flare will always set values parsed to the CLI itself first, if no value was parsed it will check if the configuration YAML contains this setting and if neither of these methods are fulfilled, \flare sets a default value as per \cref{tab:cli-information}. It also possible to use a mixture of CLI options and configuration YAML options, following the hierarchy detailed previously. As an example if the configuration YAML contains a value for the \texttt{name} option e.g

\begin{lstlisting}[style=plain]
# config.yaml
name: my_analysis\end{lstlisting}

and a value is parsed to the \texttt{--name} option of the \flare CLI.

\begin{lstlisting}[style=plain]
$ ls
stage1.py config.yaml
$ flare run analysis --name their_analysis\end{lstlisting}
Then \flare will set \textit{their\_analysis} as the name of the analysis being conducted. 

\subsubsection{Running the FCC Analysis Workflow \label{subsubsec:running-fcc-workflow}}
After correctly setting up the FCC Analysis stage scripts as per \Cref{subsec:Analysis} and understanding the configuration options that the \flare CLI has, as discussed in \Cref{subsec:cli}, one can minimally run their FCC analysis by executing the following command in the terminal: 

\begin{lstlisting}[style=plain]
$ flare run analysis\end{lstlisting}

\subsubsection{Running the MC Production Workflow}
After correctly setting up the MC Production input files required for your generator of choice (as per \Cref{subsec:mcproduction-config}) and understanding the configuration options that the \flare CLI has in \Cref{subsec:cli}, one can minimally run the MC Production workflow by executing the following command in the terminal:

\begin{lstlisting}[style=plain]
$ flare run mcproduction
\end{lstlisting}

\subsubsection{MC Production and Analysis Workflow \label{subsubsec:running-mc-prod}}
\flare is able to connect the MC Production Workflow and FCC Analysis Workflow together. An analyst must ensure all required input files are present as per \Cref{subsec:Analysis} and \Cref{subsec:mcproduction-config} and any additional configuration is set as per \Cref{subsec:cli}, one can execute the following command in the terminal to begin the workflow:

\begin{lstlisting}
$ flare run analysis --mcprod\end{lstlisting}

By adding the \texttt{--mcprod} option \flare will schedule the MC production Workflow as the inputs of the FCC Workflow. The ordering is ensured by b2luigi such that all the MC will be generated prior to running the FCC Analysis portion of the workflow. 

\subsection{Custom Workflows in \flare (No CLI) \label{subsec:custom-flare-workflow}}

Although the \flare CLI is a helpful tool, it is also possible to run a custom workflow containing whole or parts of the \flare workflow without invoking the \flare CLI. To do this, a user would create their own python script and and import the \flare package. The \flare module contains its own versions of the \texttt{b2luigi} classes \texttt{Task}, \texttt{DispatchableTask} and \texttt{WrapperTask}. Using these classes, one can create their own \texttt{b2luigi} workflow. It is important to note that one \underline{must} use the \flare version of the \texttt{b2luigi} tasks as \flare must perform some operations on these \texttt{b2luigi} classes. For ease of use, the \flare package also gives users access to the \texttt{b2luigi} settings manager through the \texttt{get\_setting} and \texttt{set\_setting} functions. For more details about these \texttt{b2luigi} classes and setting manager, consult the \texttt{b2luigi} documentation \cite{b2luigi2025}. To access all of these \flare features, one must import them at the beginning of their python script.

\begin{lstlisting}
from flare import {
    get_setting,
    set_setting,
    Task,
    DispatchableTask,
    WrapperTask
}
\end{lstlisting}

\flare offers well rounded customization options for users wanting to create their own custom workflow inside \flare. With options to connect an existing \flare Task or entire \flare Workflow into a custom workflow. To run a custom workflow using \flare one must use the \texttt{process} function, this is the entry point for all \flare related workflows, for example:

\begin{lstlisting}
# my_script.py
import flare

class MyTask(flare.Task):
    # details of the task

if __name__ == "__main__":
   flare.process(
    MyTask()
   )
\end{lstlisting}

Running the workflow works as any python script does, by calling the \texttt{python3} CLI tool.

\begin{lstlisting}
$ python3 my_script.py
\end{lstlisting}

One feature of running \flare in this way is that all the \flare CLI options discussed inside \Cref{subsec:cli} are still available on the command line. The only additional requirement of the user is to use the \texttt{get\_args} function from the \flare module to access these options, for example:

\begin{lstlisting}
# my_script.py
import flare

class MyTask(flare.Task):
    # details of the task

if __name__ == "__main__":
   flare.process(
    MyTask(),
    flare_args = flare.get_args()
   )
\end{lstlisting}
Then one can use any option available in the normal \flare CLI
\begin{lstlisting}
$ python3 my_script.py --name MyAnalysis
\end{lstlisting}

One can also just declare the CLI option values inside the python script by interacting with the object returned by the \texttt{get\_args()} function as shown below.

\begin{lstlisting}
# my_script.py
import flare

class MyTask(flare.Task):
    # details of the task

if __name__ == "__main__":
   args =  flare.get_args()
   args.name = "MyAnalysis"
   
   flare.process(
    MyTask(),
    flare_args = args
   )
\end{lstlisting}

In addition to all these features, the \flare module also gives access to all the different Workflow tasks discussed in \Cref{subsec:Analysis} and \Cref{subsec:mcproduction-config}. To access these tasks, one must use the functions \texttt{get\_mc\_prod\_stages\_dict} and \texttt{get\_fcc\_stages\_dict}. Since the exact tasks to be ran are not known until runtime, these functions will return ordered dictionaries with the required tasks. If for example a user wished to create workflow in which they download the Whizard .sin files and then ran the \flare MC Production workflow, they could do something like this:

\begin{lstlisting}
import flare

class DownloadWhizardSin(flare.Task):
    # code to download whizard file

if __name__ == "__main__":
    # Make the Stage 1 task require the DownloadWhizardSin task
    mcprod_tasks = flare.get_mc_prod_task_dict(
      inject_stage1_dependency=DownloadWhizardSin
    )
    # Get the last task of the ordered dict
    last_task = next(reversed(mcprod_tasks.values()))
    # Process the workflow
    flare.process(
    last_task()
    )
\end{lstlisting}

The \texttt{inject\_stage1\_dependency} parameter is present in both \texttt{get\_mc\_prod\_task\_dict} and \texttt{get\_fcc\_stages\_dict}. It takes another \flare Task as a parameter and is one of the most useful features of \flare as it can connect the existing Workflows inside \flare to another \flare Task or even an entirely separate \flare Workflow. Using these building blocks one can in theory build any custom workflow they wish with any of the tasks inside of \flare. This design style means \flare is entirely modular in which different Workflow components can be connected together. 

Using all these features in conjunction means a user of \flare can build any custom Workflow they wish to suit their needs in an easy modular fashion

\section{\flare Examples \label{sec:flare-examples}}
In this section we will show a variety of examples using \flare, displaying its ease of use and far reaching capabilities. The code for each example can be found in the following github repository  \href{https://github.com/CamCoop1/FLARE-examples}{Flare Examples}\footnote{Examples github: https://github.com/CamCoop1/FLARE-examples}

\subsection{FCC Analysis: Higgs Mass \label{sec:higgs-mass}}
The natural first example for \flare is to take an existing example from the FCCAnalyses package and recreate the results. Using the example from \href{https://github.com/HEP-FCC/FCCAnalyses/tree/master/examples/FCCee/higgs/mH-recoil}{{FCCee higgs mH-recoil example}}\footnote{FCCee higgs mH-recoil example: https://github.com/HEP-FCC/FCCAnalyses/tree/master/examples/FCCee/higgs/mH-recoil} we begin by downloading the corresponding input MC files from \href{https://fccsw.web.cern.ch/tutorials/apr2023/tutorial2/}{FCC Tutorials April 2023}\footnote{https://fccsw.web.cern.ch/tutorials/apr2023/tutorial2/}. The example calls for the signal sample \texttt{p8\_ee\_ZH\_Zmumu\_ecm240} and background sample \texttt{p8\_ee\_ZZ\_mumubb\_ecm240} MC, to be downloaded with a simple \texttt{wget} command. 

\begin{figure}
    \centering
    \includegraphics[width=0.8\linewidth]{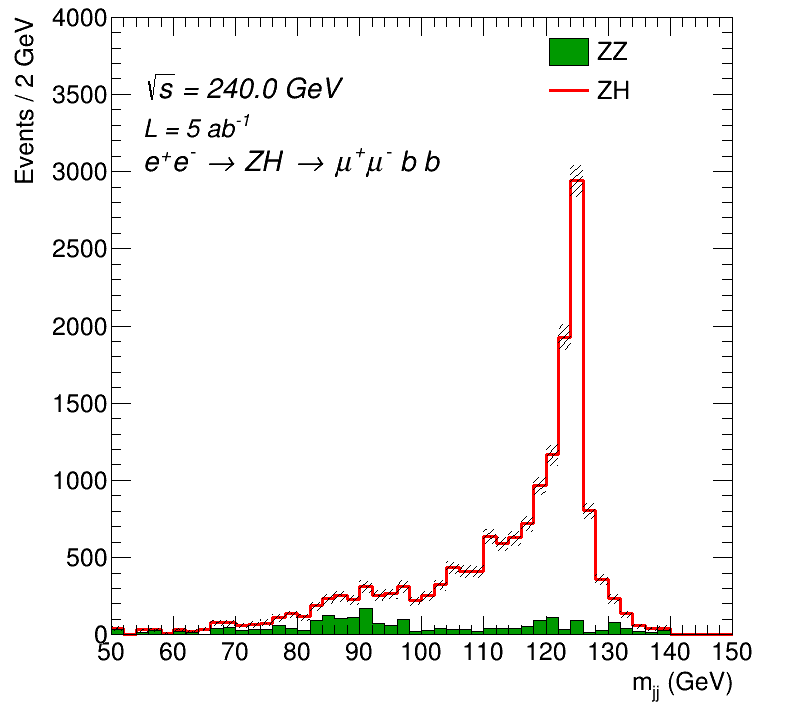}
    \caption{Reconstructed two jet mass for signal $e^+e^- \to ZH \to \mu^+\mu^- bb$  and background $e^+e^- \to ZZ \to \mu^+\mu^- bb$}
    \label{fig:higgs-example:jjmass}
\end{figure}

\begin{figure}
    \centering
    \includegraphics[width=0.8\linewidth]{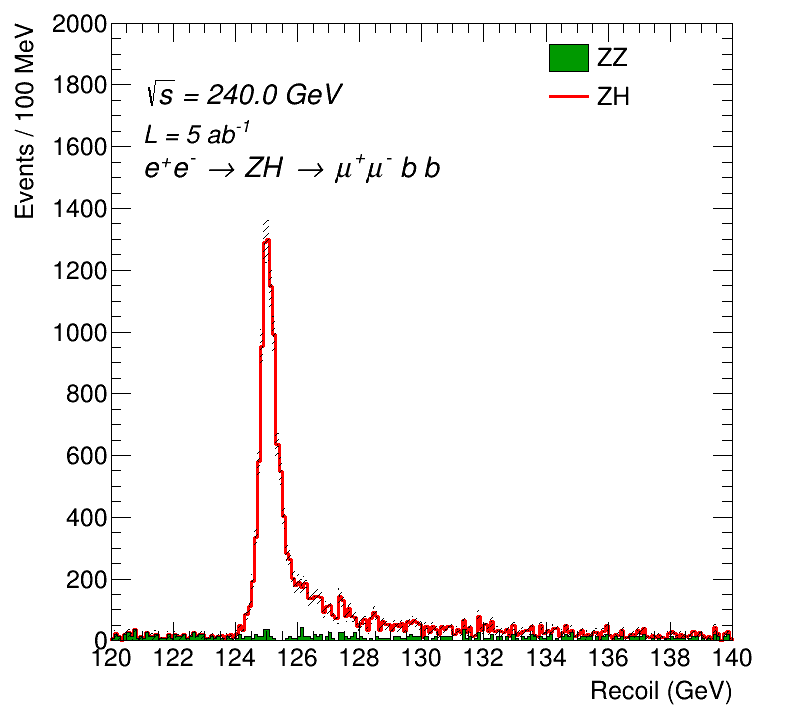}
    \caption{Reconstructed recoil mass for signal $e^+e^- \to ZH \to \mu^+\mu^- bb$  and background $e^+e^- \to ZZ \to \mu^+\mu^- bb$}
    \label{fig:higgs-example:zmumu-recoil-mass}
\end{figure}

We establish the FCC Analysis Workflow required for this example as per \Cref{subsec:Analysis}. Reviewing the example repository we find there to be a Stage1, Stage2 and Plots stages, requiring three python scripts each uniquely prefixed with their corresponding stage. Running this workflow as per \Cref{subsubsec:running-fcc-workflow}, \Cref{fig:higgs-example:jjmass} and \Cref{fig:higgs-example:zmumu-recoil-mass} show the outputs of this analysis. 

\subsection{Large Scale MC Production Time Comparison \label{sec:example-large-batch-time}}
The next example displays the power of \flare and its MC Production Workflow. We will conduct timed experiments for the MC Production configuration YAML shown below in which we will generate four MC samples using the Whizard MC Production Workflow.

\begin{lstlisting}
'$model' : UserMCProdConfigModel

global_prodtype : whizard

datatype:
    - wzp6_ee_nunuH_Hbb_ecm240
    - wzp6_ee_mumuH_Hbb_ecm240
    - wzp6_ee_bbH_HWW_ecm240
    - wzp6_ee_bbH_Hbb_ecm240\end{lstlisting}

To conduct our timed experiments, we submit four seperate \flare MC Production Workflows in parallel and time each one, taking the average as our final result. We repeat this experiment for generating 1000 and 10,000 events. Table \ref{tab:large-batch-system-1000} shows the results for 1000 events with an average time of 247.6 seconds, or 04:07. Table \ref{tab:large-batch-system-10000} shows the results for the 10,000 event test with an average time of 964.6 seconds or 16:04, only 4 times longer for 10 times more events. 

This example displays the ease with which one can dispatch any number of MC production workflows using \flare with no additional configuration code from the analyst. Due to \flare and its Workflows the two set Whizard production process is made easy, requiring a single execution from the user.

\begin{table}[h]
    \centering
    \caption{Large batch timing example in seconds, 1000 events}
    \begin{tabular}{c|c}
    \textbf{Nth Trial} & \textbf{Time (sec)} \\ \hline
    1 & 262.6 \\
    2 & 242.5 \\
    3 & 242.5 \\
    4 & 242.5 \\ \hline 
    Average & 247.6 
    \end{tabular}
    \label{tab:large-batch-system-1000}
\end{table}

\begin{table}[h]
    \centering
    \caption{Large batch timing example in seconds, 10,000 events}
    \begin{tabular}{c|c}
    \textbf{Nth Trial} & \textbf{Time (sec)} \\ \hline
1 & 964.6 \\
2 & 1044.83\\
3 & 904.5 \\
4 & 944.5 \\ \hline
Average & 964.6
    \end{tabular}
    \label{tab:large-batch-system-10000}
\end{table}

\subsection{Whizard Cross Section Comparison \label{sec:example-whizard-crossection}} 
One powerful part of \flare is its ability to be flexible and integrate into any \texttt{b2luigi} workflow created by a user. This example aims to show an instance where a user can build their own \texttt{b2luigi} workflow containing the internal \flare tasks to conduct an analysis. The specifics of building a custom Workflow like this this are discussed inside \Cref{subsec:custom-flare-workflow}.

The goal of this example is to calculate the cross section of $e^+e^- \to HX\bar{X} \to Y\bar{Y}X\bar{X}$ where X and Y are placeholders for potential decay products. To do this we use the cross section calculated by Whizard for the $e^+e^-\to HX\bar{X}$ portion of the sample decay along with the PDG branching fractions for Higgs decay $H\to Y\bar{Y}$ provided by the \href{https://pdg.lbl.gov/2024/reviews/rpp2024-rev-higgs-boson.pdf}{PDG Higgs Branching Fractions}\footnote{Higgs Decays: https://pdg.lbl.gov/2024/reviews/rpp2024-rev-higgs-boson.pdf}, to calculate the overall cross section of the produced MC sample. This is represented by \Cref{eq:cross-section}.

\begin{equation}
    \label{eq:cross-section}
    \sigma_{total} = \sigma_{e^+e^-\to HX\bar{X}} \times \mathcal{BF}(H\to Y\bar{Y})
\end{equation}

We will then compare our calculated cross section to that of the centrally produced MC for the FCC Collaboration located at \href{https://fcc-physics-events.web.cern.ch/fcc-ee/delphes/winter2023/idea/}{fcc-ee winter2023 MC}\footnote{https://fcc-physics-events.web.cern.ch/fcc-ee/delphes/winter2023/idea/}.

\smallbreak
The exact implementation of this example can be found here \href{https://github.com/CamCoop1/FLARE-examples/tree/main/analysis/studies/calculate_whizard_cross_section_example}{Calculate Whizard Cross Section}\footnote{https://github.com/CamCoop1/FLARE-examples/tree/main/analysis/studies/}. This example makes use of the techniques discussed in \Cref{subsec:custom-flare-workflow}, in which we will run this workflow without invoking the \flare CLI.

\smallbreak

In this example we will take the Stage1 Whizard Task, named \texttt{MCProductionStage1}, from \texttt{flare.get\_mc\_prod\_stages\_dict} and build a workflow that takes the log file produced from Whizard and calculate the cross section for each input datatype and then compiles these cross sections into a single output file. The MC types we will be producing are as follows:

\begin{itemize}
    \item \texttt{wzp6\_ee\_nunuH\_Hbb\_ecm240}
    \item \texttt{wzp6\_ee\_mumuH\_Hbb\_ecm240}
    \item \texttt{wzp6\_ee\_bbH\_HWW\_ecm240}
    \item \texttt{wzp6\_ee\_bbH\_Hbb\_ecm240}
\end{itemize}

\begin{figure*}
    \centering
    \includegraphics[width=\linewidth, height=7cm]{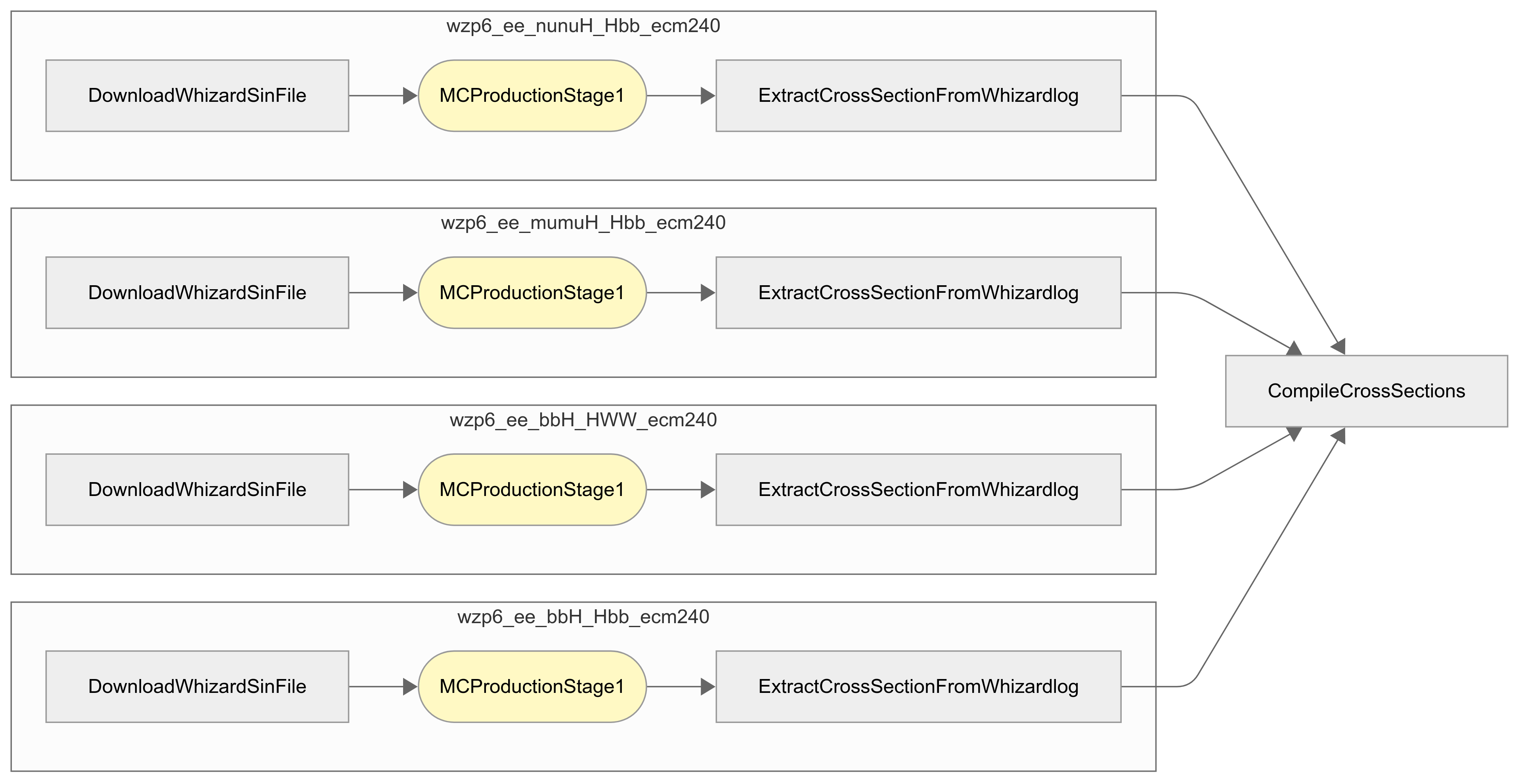}
    \caption{Mermaid graph showing the custom workflow used to generate the example in \Cref{sec:example-whizard-crossection}. Note that light yellow MCProductionStage1 Tasks are indicating that task is from the internal FLARE workflow whilst the other tasks are custom made. Each datatype has its own linear workflow that is ran in parallel during runtime. The outputs of the ExtractCrossSection Task is then collected into a single output produced by CompileCrossSections. This chart was produced using the Mermaid-chart web plugin \cite{mermaidchart2025}}
    \label{fig:whizard-cross-section-mermaid}
\end{figure*}

The workflow is detailed in the mermaid chart shown in Figure \ref{fig:whizard-cross-section-mermaid}. Note that there is a linear workflow for each datatype which includes the MCProductionStage1 Task from inside \flare. All of these liner workflows are then compiled into a singular output via the CompileCrossSections Task. 

The MC production config file used for this example is shown below.

\begin{lstlisting}[language=yaml, style=yamlstyle]
'$model' : UserMCProdConfigModel

global_prodtype : whizard

datatype:
    - wzp6_ee_nunuH_Hbb_ecm240
    - wzp6_ee_mumuH_Hbb_ecm240
    - wzp6_ee_bbH_HWW_ecm240
    - wzp6_ee_bbH_Hbb_ecm240\end{lstlisting}

\begin{figure}[h]
    \centering
    \includegraphics[width=0.8\linewidth]{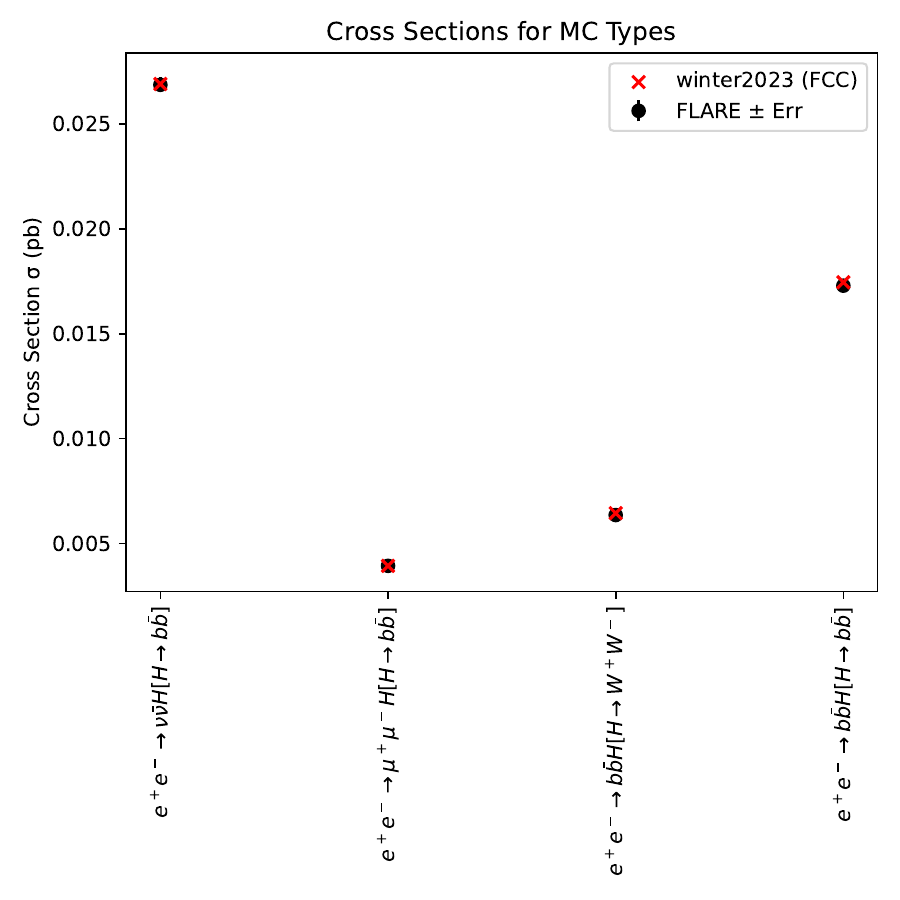}
    \caption{Visual display of results shown in \Cref{tab:whizard-cross-sections}. The black dots indicates the calculated cross section in this example for the given decay (x-axis) along with its error. The red crosses indicate the cross section of the centrally produced MC samples.}
    \label{fig:whizard-cross-section-scatter}
\end{figure}

The results are shown in Table \ref{tab:whizard-cross-sections}. All calculated cross sections are within uncertainties of their centrally produced value, as expected. A visual representation of this is shown in \Cref{fig:whizard-cross-section-scatter}

\begin{table}[h]
    \centering
    \caption{Calculated cross section using Whizards output compared to centrally produced cross sections at \href{https://fcc-physics-events.web.cern.ch/fcc-ee/delphes/winter2023/idea/}{\textcolor{blue}{fcc-ee winter2023 MC}}. Here we use the python package \textit{uncertainties} \cite{uncertanties2025} to propagate the uncertainties from the Higgs branching fractions and the cross section calculated from Whizard. Note that all cross section calculations from this example are within uncertainties of their reference values from the centrally produced samples at  \href{https://fcc-physics-events.web.cern.ch/fcc-ee/delphes/winter2023/idea/}{fcc-ee winter2023 MC} } 

    \begin{tabular}{l|c|c}
    \textbf{MC Type} & Calc $\sigma$ (pb) & Central $\sigma$ (pb) \\ \hline
$e^+e^-\to\nu\bar{\nu} H[H\to b\bar{b}]$& 0.02686$\pm$0.00035 & 0.0269  \\
$e^+e^-\to\mu^+\mu^- H[H\to b\bar{b}]$ & 0.00394$\pm$0.00005 & 0.00394 \\
$e^+e^-\to b\bar{b}H[H\to W^+W^-]$ &  0.00636$\pm$0.00010 & 0.00645\\
$e^+e^-\to b\bar{b}H[H\to b\bar{b}]$ & 0.01729$\pm$0.00022 & 0.01745 \\

    \end{tabular}
    \label{tab:whizard-cross-sections}
\end{table}

\subsection{Multi-Detector MC Production to FCC Fast Simulation Higgs Study \label{sec:example-full-fastsim}}

\begin{figure*}
    \centering
    \includegraphics[width=0.99\linewidth]{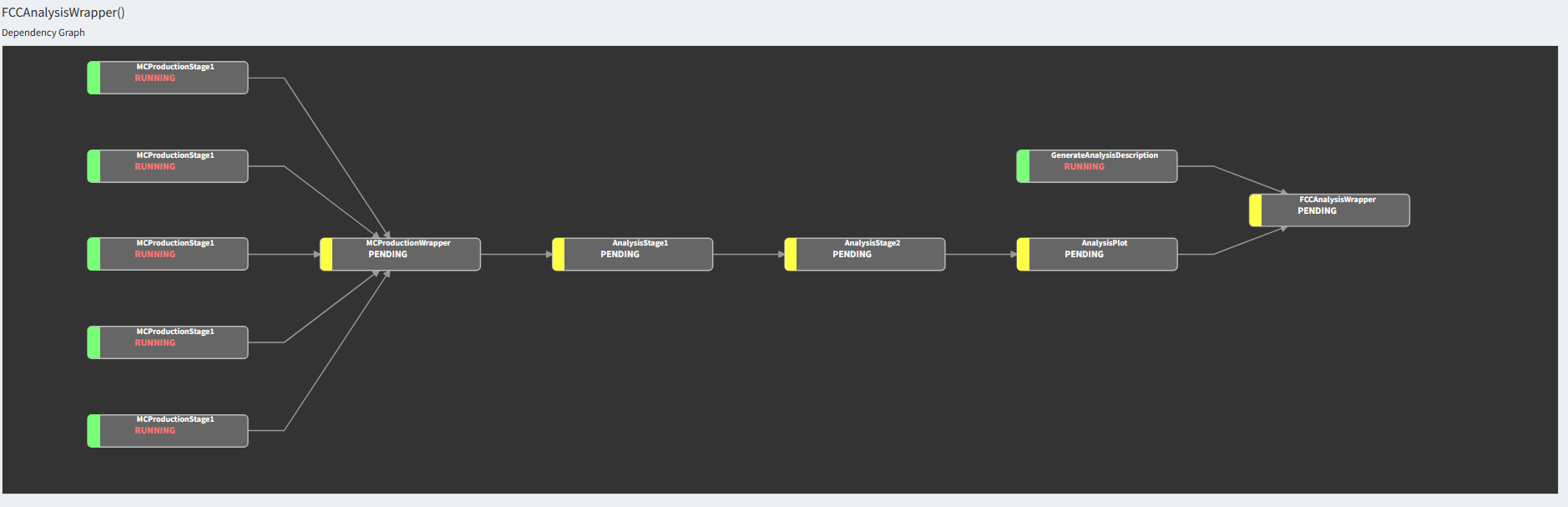}
    \caption{luigid real-time DAG graph showing the full MC Production - FCC analysis workflow conducted using \flare}
    \label{fig:fastsime-workflow}
\end{figure*}
In this example we will bring many of the features of \flare to the forefront as we conduct the Fast Simulation Higgs mass study from \Cref{sec:higgs-mass} looking at different detector configurations. Using five different detector cards and a single decay type $e^+e^-\to ZH \to \mu^+\mu^-H$, we will generate five different MC root files each containing 50,000 events via the \flare MC Production workflow. These root files will be analysed by the FCC analysis Workflow. An identical Stage1 from \Cref{sec:higgs-mass} is used but an altered Stage2 and Plot scripts with select variables and adjustments to overlay each detector type in each plot. The MC configuration YAML is shown below, something to note is the need for the \texttt{global\_env\_script\_path}, this is required because this Fast Simulation analysis example is only compatable with MC produced using an older version of the \texttt{Key4HEP} stack. The details of how this environment setup script is used can be found in \Cref{subsubsec:specific-env}. 

\begin{lstlisting}
"$model" : "UserMCProdConfigModel"

global_prodtype: pythia8

global_env_script_path : setup.sh

datatype:
    - p8_ee_ZH_Zmumu_ecm240 

card:
    - card_IDEA
    - card_IDEA_lighterVXD_35pc
    - card_IDEA_lighterVXD_50pc
    - card_IDEA_3T
    - card_IDEA_SiTracking\end{lstlisting}
    
During the running of this workflow, we used the \texttt{luigid} commandline tool which is apart of the luigi package. This allows us to display, monitor and edit any \texttt{b2luigi} workflow during runtime. \Cref{fig:fastsime-workflow} shows an example of the displayed workflow, we note that there are five \texttt{MCProductionStage1} Tasks for each combination of \texttt{datatype} and \texttt{card}. These are combined into a single directory by the \texttt{MCProductionWrapper} Task. The directory is then dynamically set as the \texttt{inputDir} variable inside our \texttt{AnalysisStage1} task by \flare, no adjustments required by the analyst.

\begin{figure}
    \centering
    \includegraphics[width=0.72\linewidth]{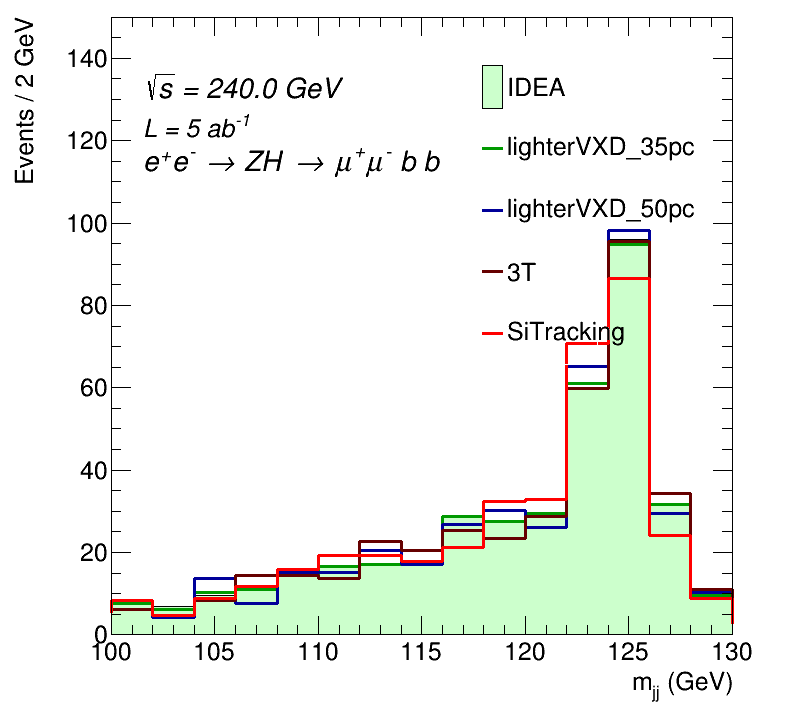}
    \caption{Combined jet mass for $ee\to Z[Z\to\mu\mu]H$ for each detector card}
    \label{fig:fastsim-example:jjmass}
\end{figure}

\begin{figure}
    \centering
    \includegraphics[width=0.72\linewidth]{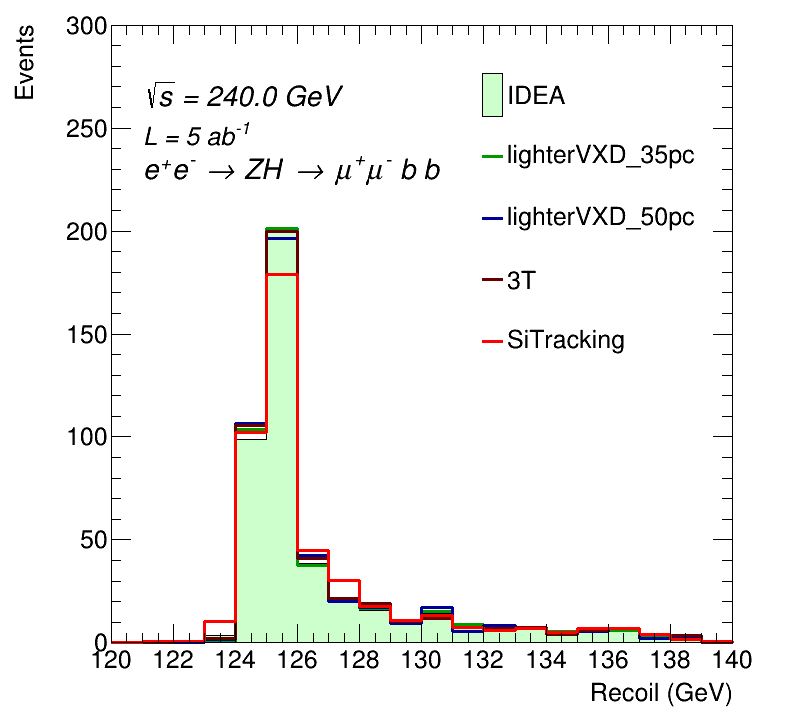}
    \caption{Recoil mass for $ee\to Z[Z\to\mu\mu]H$ for each detector card}
    \label{fig:fastsim-example:zmumu-recoil-mass}
\end{figure}

The Plot stage script was altered such that only the recoil mass and combined jet mass histograms are created. \Cref{fig:fastsim-example:jjmass} shows the combined jet mass distributions for each detector card, similarly \Cref{fig:fastsim-example:zmumu-recoil-mass} shows the recoil mass. Since this is just an example to display the feature rich \flare, we will not delve into the histograms or what they may be conveying.

\section{Conclusions}
This paper serves as a detailed overview of \flare discussing the package and its features along with examples of its use. \Cref{sec:flare-package} discussed \texttt{b2luigi} and its basic concepts, installing \flare and the configuration required to run the FCC Analysis and MC Production Workflows. The connivent \flare CLI was detailed in \Cref{subsec:cli} and its various options. Also discussed in \Cref{subsec:custom-flare-workflow} is the features inside the \flare package making is possible for a user to create their own custom Workflows using parts of whole \flare Workflows. All of the teachings of \Cref{sec:flare-package} are then put on full display inside \Cref{sec:flare-examples} where we go through four different examples using \flare. \flare is open source and anyone interested in contributing is encouraged to do so at \href{https://github.com/CamCoop1/FLARE}{https://github.com/CamCoop1/FLARE}

\section{Acknowledgments}
This software would not have been created if not for the many conversations and input from colleagues in the Physics department of the University of Adelaide. It was because of these conversations and many iterations that \flare is as feature rich as it is in this version 0.1.4. This collaborative environment is at the core of \flare and its ethos, open source and with extensibility in mind.

\section{Useful Links}
\begin{itemize}
    \item[] \href{https://github.com/CamCoop1/FLARE}{https://github.com/CamCoop1/FLARE}
    \item[] \href{https://github.com/CamCoop1/FLARE-examples/}{https://github.com/CamCoop1/FLARE-examples/}
    \item[] \href{https://pypi.org/project/hep-flare/}{https://pypi.org/project/hep-flare/}
    \item[] \href{https://zenodo.org/records/15694669}{https://zenodo.org/records/15694669}
    \item[] \href{https://b2luigi.belle2.org/}{https://b2luigi.belle2.org/}
    \item[] \href{https://hep-fcc.github.io/FCCAnalyses/}{https://hep-fcc.github.io/FCCAnalyses/}
    \item[] \href{https://key4hep.github.io/key4hep-doc/}{https://key4hep.github.io/key4hep-doc/}
    
\end{itemize}

\bibliography{main}

\end{document}

%% file: code_stylings.tex
\usepackage{listings}
\usepackage{xcolor}

\definecolor{codegreen}{rgb}{0,0.6,0}
\definecolor{codegray}{rgb}{0.5,0.5,0.5}
\definecolor{codepurple}{rgb}{0.58,0,0.82}
\definecolor{backcolour}{rgb}{0.95,0.95,0.92}

\definecolor{codepurple}{rgb}{0.58,0,0.82}
\definecolor{backcolour}{rgb}{0.95,0.95,0.92}
\definecolor{keywordcolor}{rgb}{0.36,0.54,0.66}
\definecolor{commentcolor}{rgb}{0.4,0.4,0.4}

\lstdefinelanguage{yaml}{
  morekeywords={true,false,null,yes,no},
  sensitive=true,
  morecomment=[l]{\#},
  morestring=[b]',
  morestring=[b]",
  alsoletter={:},
  keywordstyle=\color{keywordcolor}\bfseries,
  commentstyle=\color{commentcolor}\itshape,
  stringstyle=\color{codepurple},
}

\lstdefinestyle{yamlstyle}{
  backgroundcolor=\color{backcolour},
  basicstyle=\ttfamily\small,
  breaklines=true,
  showstringspaces=false,
  columns=fullflexible
}

\lstdefinestyle{pythonstyle}{
    backgroundcolor=\color{backcolour},   
    commentstyle=\color{codegreen},
    keywordstyle=\color{magenta},
    numberstyle=\tiny\color{codegray},
    stringstyle=\color{codepurple},
    basicstyle=\ttfamily\footnotesize,
    breakatwhitespace=false,         
    breaklines=true,                 
    captionpos=b,                    
    keepspaces=true,                 
    numbers=left,                    
    numbersep=5pt,                  
    showspaces=false,                
    showstringspaces=false,
    showtabs=false,                  
    tabsize=2
}

\lstdefinestyle{plain}{
  basicstyle=\ttfamily \footnotesize,     
  keywordstyle=,
  commentstyle=,
  stringstyle=,
  showstringspaces=false,
  breaklines=true,
  language=,
}